%% file: main.tex
\documentclass{IEEEtran}
\usepackage{cite}
\ifCLASSINFOpdf
   \usepackage[pdftex]{graphicx}
\else
   \usepackage[dvips]{graphicx}
\fi
\usepackage[cmex10]{amsmath}
\usepackage{amssymb}
\usepackage[caption=false,font=footnotesize]{subfig}

\usepackage{comment}

\usepackage{tabularx,booktabs,makecell,multirow,array}
\usepackage[hidelinks]{hyperref}
\usepackage{algorithm}
\usepackage{algpseudocode}

\makeatletter
\let\old@ps@headings\ps@headings
\let\old@ps@IEEEtitlepagestyle\ps@IEEEtitlepagestyle
\def\footer#1{%
    \def\ps@headings{%
        \old@ps@headings%
        \def\@oddfoot{\strut\hfill#1\hfill\strut}%
        \def\@evenfoot{\strut\hfill#1\hfill\strut}%
    }%
    \def\ps@IEEEtitlepagestyle{%
        \old@ps@IEEEtitlepagestyle%
        \def\@oddfoot{\strut\hfill#1\hfill\strut}%
        \def\@evenfoot{\strut\hfill#1\hfill\strut}%
    }%
    \ps@headings%
}
\makeatother

\footer{%
        This preprint is currently under peer review. 
}

\begin{document}
\title{Addressing Model Inaccuracies in Transmission Network Reconfiguration via Diverse Alternatives}

%% To specify the authors when (number of affiliations <= 2)
\author{
\IEEEauthorblockN{Paul Bannmüller\IEEEauthorrefmark{1}, Périne Cunat\IEEEauthorrefmark{1} \IEEEauthorrefmark{7}, Ali Rajaei\IEEEauthorrefmark{1}, Jochen Cremer\IEEEauthorrefmark{1} \IEEEauthorrefmark{7}}
\IEEEauthorblockA{\IEEEauthorrefmark{1} Department of Electrical Sustainable Energy, Delft University of Technology, Delft, The Netherlands\\
\IEEEauthorrefmark{7}AIT Austrian Institute of Technology GmbH, Vienna, Austria\\
\{p.e.bannmuller, p.e.a.cunat, a.rajaei, j.l.cremer\}@tudelft.nl}

}

%% To specify the authors when (number of affiliations > 2)
% \author{\IEEEauthorblockN{Author n.1\IEEEauthorrefmark{1},
% Author n.2\IEEEauthorrefmark{2},
% Author n.3\IEEEauthorrefmark{3}, 
% Author n.4\IEEEauthorrefmark{3} and
% Author n.5\IEEEauthorrefmark{4}}
% \IEEEauthorblockA{\IEEEauthorrefmark{1} Department Name of Organization A\\
% Name of the organization A,
% Address A\\ Emails if wanted}
% \IEEEauthorblockA{\IEEEauthorrefmark{2} Department Name of Organization B\\
% Name of the organization B,
% Address B\\ Emails if wanted}
% \IEEEauthorblockA{\IEEEauthorrefmark{3} Department Name of Organization C\\
% Name of the organization C,
% Address C\\ Emails if wanted}
% \IEEEauthorblockA{\IEEEauthorrefmark{4}Department Name of Organization D\\
% Name of the organization D,
% Address D\\ Emails if wanted}
% }

\maketitle

\begin{abstract}
%The Latex template and basic guidelines for the preparation of a technical paper for the PSCC 2026 conference are presented. The abstract is limited to 150 words and cannot contain equations, figures, tables, or references. It should concisely state what was done, how it was done, principal results, and their significance.
\input{abstract}
\end{abstract}


% Use this to place sponsorships
\thanksto{\noindent The work of P. Bannmüller and J.L. Cremer, are supported through the AI-EFFECT project (Grant Agreement No 101172952) funded under the European Union's Horizon Europe Research and Innovation program. However, views and opinions expressed are those of the author(s) only and do not necessarily reflect those of the European Union. Neither the European Union nor the granting authority can be held responsible.}

\input{introduction}
\input{methodology}

\input{case_studies}
\input{conclusion}

% The 'triggered' command can be changed if desired:


%\bibliographystyle{IEEEtran}
%\newpage
\bibliographystyle{IEEEtran}
\bibliography{Bib_complete}

\end{document}

%% file: abstract.tex
The ongoing energy transition places significant pressure on the transmission network due to increasing shares of renewables and electrification. To mitigate grid congestion, transmission system operators need decision support tools to suggest remedial actions, such as transmission network reconfigurations or redispatch. However, these tools are prone to model inaccuracies and may not provide relevant suggestions with regard to important unmodeled constraints or operator preferences. We propose a human-in-the-loop modeling-to-generate alternatives (HITL-MGA) approach to address these shortcomings by generating diverse topology reconfiguration alternatives. Case studies on the IEEE 57-bus and IEEE 118-bus systems show the method can leverage expert feedback and improve the quality of the suggested topology reconfigurations.

%% file: introduction.tex
\section{Introduction}

The operation of the power grid plays a crucial role in enabling the energy transition. 
As the electrification of the energy demand and the share of decentralized variable renewable energy sources in our energy supply increase, more pressure is placed on the transmission system, regularly leading to grid congestion, as in the Netherlands~\cite{ieaNetherlands20242025}. 
Reconfiguring the topology of the transmission grid as a remedial action can mitigate grid congestion cost-effectively. Such topology reconfiguration increases the flexibility of the transmission grid by switching lines, splitting busbars, or reconfiguring substations~\cite{hanCongestionManagementTopological2015}. 

The combinatorial complexity of the topology reconfiguration problem poses a major challenge for the transmission system operators and the necessary decision support tools, even if simplified modeling assumptions, such as linearized power flow (DC-PF), are applied \cite{heidarifarNetworkTopologyOptimization2016, kocukNewFormulationStrong2017, pinedaTightBigMsOptimal2024}. 
Modeling details that are necessary to apply topology reconfiguration in power system operation, like system stability or the aleatoric uncertainty in load, generation, and contingencies, further increase complexity \cite{khanabadiOptimalTransmissionSwitching2013, hanOptimalTransmissionSwitching2023, hanBumplessTopologyTransition2023}. Consequently, approaches proposed in the literature leveraging techniques of reduction and decomposition \cite{wangPowerGridDecomposition2018, sarOptimizingPowerGrid2025} fail to decrease complexity sufficiently to include all necessary modeling details. 
Further approaches apply various forms of learning to reduce computational complexity while maintaining a certain level of model accuracy. These approaches aim to learn, for example, parts of optimization models and policies or to switch actions directly \cite{pinedaLearningassistedOptimizationTransmission2024, dorferPowerGridCongestion2022, lautenbacherMultiobjectiveReinforcementLearning2025, hassounaLearningTopologyActions2025}. However, learning approaches often fail to generalize sufficiently, thus cannot fully mitigate model inaccuracies \cite{hassounaLearningTopologyActions2025, jongGeneralizableGraphNeural2025}. 

Recently, a new paradigm has emerged to face model inaccuracies: exploring a variety of alternative actions, instead of increasing the modeling detail.
For example, the GridOptions tool \cite{viebahnC2GridOptionsTool} proposes a workflow to generate a set of topological reconfiguration alternatives that satisfy different trade-offs between various objectives and are behaviorally diverse. However, the GridOptions workflow is built on a brute-force evaluation of all possible actions followed by several sequential filtering steps, including metaheuristics. The integration of different objectives at various stages of the workflow complicates the quantification of their individual impact. 
The work of \cite{dorferPowerGridCongestion2022} proposes an approach based on AlphaZero using tree search to provide different topological alternatives from which the human operator can choose. Their alternatives were generated according to the best performance with respect to a primary value function and do not explicitly aim for diversity to account for model inaccuracies.

An approach known as modeling-to-generate alternatives (MGA) has the potential to generate topological alternatives that address model inaccuracies.
The paradigm of MGA builds on the observation that complex decision-making problems often can not be fully captured by mathematical models \cite{brill_modeling_1982}. In this case, diverse near-optimal solutions may perform nearly as well as the optimum, while performing better with respect to unmodeled aspects. MGA is typically applied in long-term planning of energy capacity expansion, to provide alternative investment plans within a predefined budget \cite{decarolisUsingModelingGenerate2011}. To the authors' knowledge, the present work is the first application of MGA to network topology reconfiguration.

In particular, we propose to adapt the human-in-the-loop MGA (HITL-MGA) method introduced in \cite{lombardiHumanintheloopMGAGenerate2025}, which incorporates a feedback process to guide the exploration of alternatives and improve their performance with respect to the feedback's evaluation criteria.  
This feedback could be provided by a human operator or power system simulations that exceed the tractable level of complexity of the decision support tool. This paper simulates the feedback by using evaluation functions that rate the performance of the generated alternatives regarding three typical types of model inaccuracies: the missing integration of (i) preferences and (ii) expert knowledge of the operators due to difficulties in explicitly modeling them, and (iii) simplifying modeling assumptions.

The main contributions of this paper are:
\begin{itemize}
    \item A first application of MGA to a network topology reconfiguration problem, and insights into the specificity of this application,
    \item Adapting the HITL-MGA approach to address model inaccuracies in the topology reconfiguration optimization process with diverse alternatives.
\end{itemize}
Section \ref{sec:methodology} introduces the baseline MGA and HITL-MGA approaches, the proposed adapted HITL-MGA method, and a more detailed formulation of the considered model inaccuracies. Section \ref{sec:case_studies} presents the results of the different approaches for the IEEE 57-bus and IEEE 118-bus systems. Important insights from these case studies are discussed in Section \ref{sec:discussion}.

%% file: methodology.tex
\section{The proposed HITL-MGA approach}
\label{sec:methodology}
\begin{figure}[]
  \centering
  \includegraphics[width=0.9\linewidth]{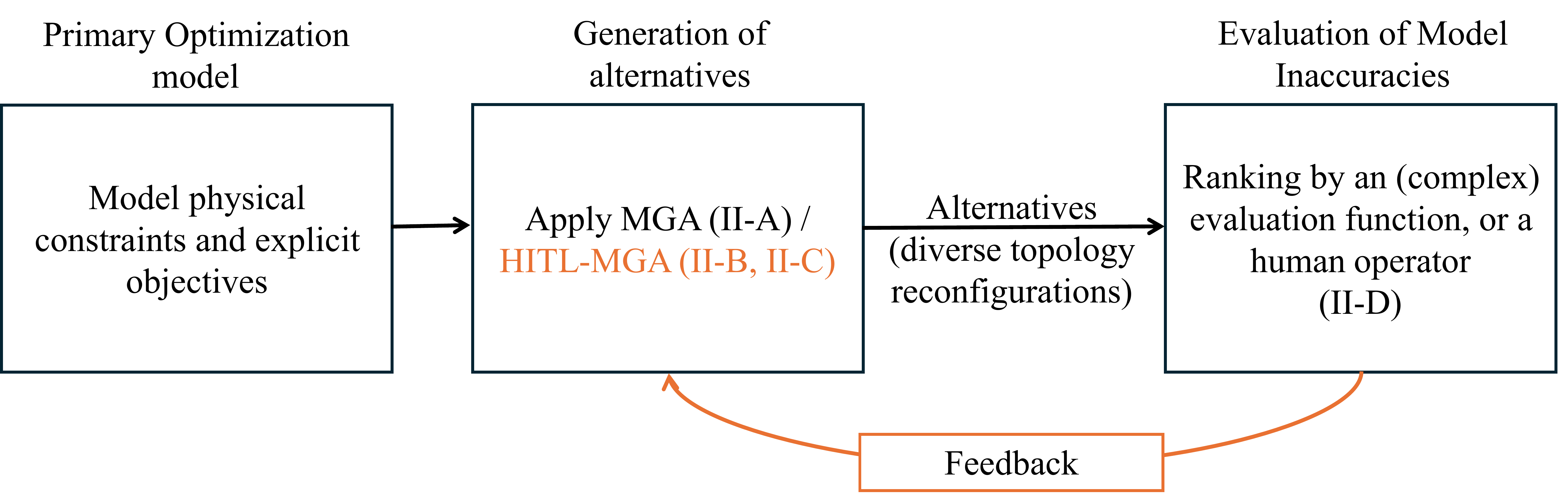}
  \caption{The proposed approach, generating initial topological alternatives with MGA, utilizing feedback from the evaluation step to guide the generation of further alternatives with HITL-MGA.}
  \label{fig:HITLMGA_overview}
\end{figure}

To guide the exploration of alternatives toward solutions that address specific model inaccuracies, we adapt the HITL-MGA approach developed by \cite{lombardiHumanintheloopMGAGenerate2025}. 
As shown in Figure~\ref{fig:HITLMGA_overview}, our approach first generates a set of alternatives via MGA (Section \ref{sec:MGA_applied_to_NTR}). The approach then computes feedback about the performance of the alternatives regarding specific model inaccuracies, in the form of a ranking of the alternatives generated by MGA (Section \ref{sec:evaluation_functions}). This feedback is used in HITL-MGA to generate new alternatives that are expected to perform better regarding the specific model inaccuracies (Sections \ref{sec:original_HITLMGA} and \ref{sec:improvements_HITL_MGA}).

\subsection{Proposed MGA Formulation for Topology Reconfiguration}
\label{sec:MGA_applied_to_NTR}
This section describes how this paper applies MGA for network topology reconfiguration based on the network reconfiguration formulation for line switching and substation reconfiguration of \cite{heidarifarNetworkTopologyOptimization2016, heidarifarOptimalTransmissionLine2021}, which poses a mixed-integer linear program (MILP) based on DC-OPF. To simplify the modeling, dispatch is modeled instead of redispatch.
We propose an augmented MGA formulation: 
\begin{subequations}
\label{eq:basic_MGA}
\begin{align}
\min_{\mathbf{x},\mathbf{y},\mathbf{z}} \quad & d_{\text{MGA}}(\mathbf{z}) = \mathbf{w}_d^T \mathbf{z} - \frac{s}{100 \cdot f^*} \label{eq:basic_MGA_objective}\\
\text{s.t.} \quad & f(\mathbf{x}) + s = f^* \cdot (1 + \varepsilon) \label{eq:basic_MGA_epsilon}\\
& \mathbf{g}(\mathbf{x,y,z}) \leq \mathbf{0} \label{eq:basic_MGA_ineq}\\
& \mathbf{h}(\mathbf{x,y,z}) = \mathbf{0} \label{eq:basic_MGA_eq}\\
& \mathbf{x} \in \mathbb{R}^m, \; \mathbf{y} \in \{0,1\}^n, \; \mathbf{z} \in \{0,1\}^q\notag
\end{align} 
\end{subequations}

\begin{figure}[]
    \centering
    \includegraphics[width=0.7\linewidth]{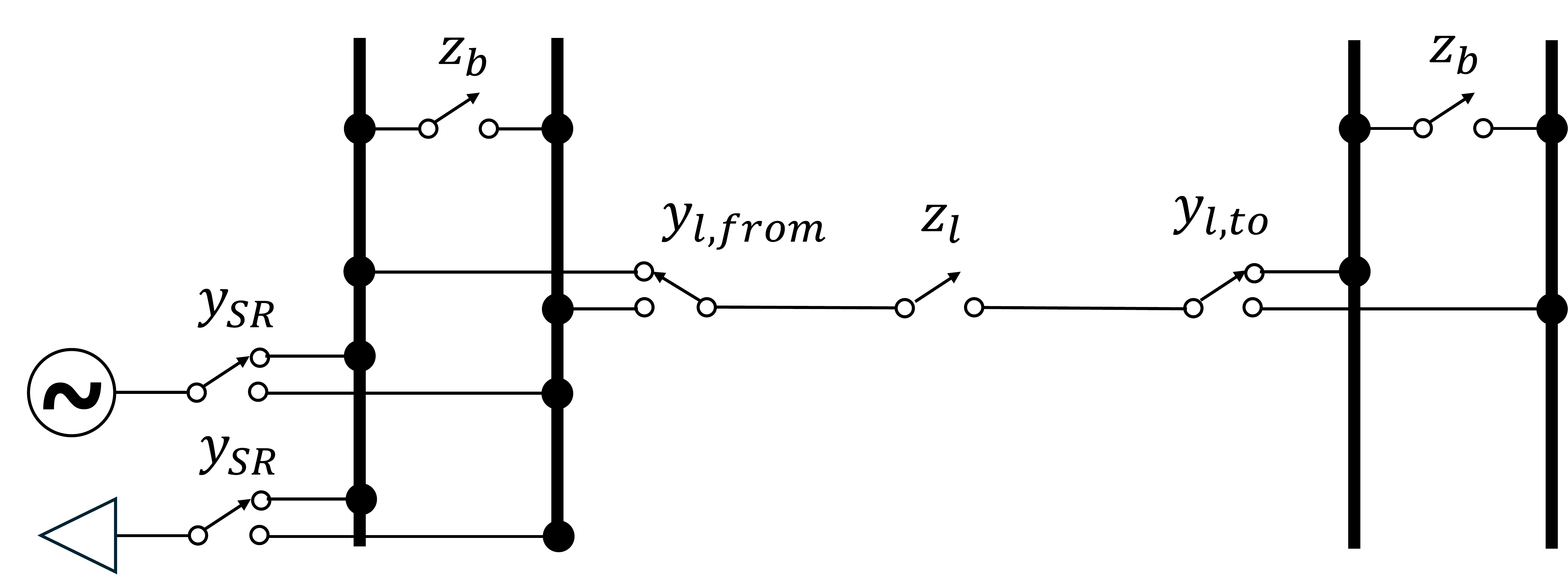}
    \caption{Modeling of substation switching, visualized for two substations with two busbars each, adapted from \cite{heidarifarNetworkTopologyOptimization2016}.}
    \label{fig:substation_switching_heidarifar}
\end{figure}

The objective models the diversity function $d_{\text{MGA}}$ as a weighted sum of the binary switching variables $\mathbf{z}$, including line switching variables $z_l$, and, in the case of substation switching, the busbar splitting variables $z_b$. 
Hence, diverse alternatives with respect to these $q$ variables in $\mathbf{z}$ can be obtained. $\mathbf{x}$ represents the dispatch variables and the other continuous operational variables (e.g., power flow, phase angle). In case of substation switching, $\mathbf{y}$ represents the binary substation reconfiguration variables $y_{\text{SR}}$ and $y_{l,\text{from/to}}$, shown in Figure \ref{fig:substation_switching_heidarifar}. The number of variables is $m$ in 
$\mathbf{x}$, and $n$ in $\mathbf{y}$.

Alternatives are obtained by solving the optimization problem \eqref{eq:basic_MGA} for different weights $\mathbf{w}_d$, which are sampled between zero and one from a uniform distribution. The $\varepsilon$-constraint \eqref{eq:basic_MGA_epsilon} ensures that all solutions remain in a near-optimal solution space, which is defined by $\varepsilon$ and the optimum of the primary objective $f^*$. This study models the primary objective $f$ as the cost of energy generation.
We augment the original MGA formulation by including the buffer variable of the primary objective $s$ in the objective function \cite{finkeLinkingModellingGenerate2024}. By normalizing the buffer variable with the optimum of the primary objective $f^*$ multiplied by a constant \eqref{eq:basic_MGA_objective}, we ensure that information about the primary objective is still present in the MGA, while on a smaller order of magnitude. This makes the generated alternative topologies optimal with respect to the MGA objective, while the dispatch $\mathbf{x}$ and the substation reconfiguration $\mathbf{y}$ are optimal with respect to the buffer variable term.
\eqref{eq:basic_MGA_ineq} and \eqref{eq:basic_MGA_eq} model the inequality and equality constraints of the topology reconfiguration problem \cite{heidarifarNetworkTopologyOptimization2016}.

\subsection{Baseline HITL-MGA Approach}
\label{sec:original_HITLMGA}
The baseline HITL-MGA approach \cite{lombardiHumanintheloopMGAGenerate2025} augments the MGA objective \eqref{eq:basic_MGA_objective} to incorporate feedback on the performance of the already generated alternatives, which is encoded in the weights $\mathbf{w}^{(k)}_{\text{HITL}}$ for each top-ranked alternative $k$. The two components of the resulting HITL-MGA objective \eqref{eq:HITL_objective} are weighted with parameters $a$ and $b$. Note that parameter $b$ scales both negative and positive weights $\mathbf{w}^{(k)}_{\text{HITL}}$, and weights are normalized.
\begin{equation}
    \label{eq:HITL_objective}
    \begin{aligned}
         d^{(k)}_{\text{HITL}}(\mathbf{z}) & = \Bigl( \underbrace{a \frac{\mathbf{w}_d^\top}{\| \mathbf{w}_d \|_2} }_A  + \underbrace{b \frac{{\mathbf{w}^{(k)}_{\text{HITL}}}^\top}{\| \mathbf{w}^{(k)}_{\text{HITL}} \|_2} }_B \Bigr) \mathbf{z} - \frac{s}{100 \cdot f^*}\\
%        w_{\text{HITL}} & \in \{-1,0,1\} \\
        a,b & \geq 0, \quad k\in K
    \end{aligned}
\end{equation}

The weights $\mathbf{w}^{(k)}_{\text{HITL}}$ are generated as expressed in \eqref{eq:HITLMGA_weights_lombardi}, where $I_{\text{MGA}}$ represents the set of all alternatives generated by MGA, $K$ represents the top-ranked alternatives in $I_{\text{MGA}}$. 
Here, for each top-ranked alternative, the difference $\delta$ between the average over all alternatives and the top-ranked alternative determines the weight of the corresponding switching action in the HITL-MGA objective. A weight $w^{(k)}_{\text{HITL},j}$ different from zero is assigned if the $\delta$ exceeds the distinctive threshold $\tau$, which is a tunable parameter. 
Furthermore, one set of weights $\mathbf{w}^{(\text{sum})}_{\text{\text{HITL}}}$ is introduced that uses the sum of all vectors of each top-ranked alternative as $\mathbf{w}_{\text{HITL}}$ weights:
\begin{subequations}
\label{eq:HITLMGA_weights_lombardi}
\begin{align}
\delta_{k,j} &= \mathbb{E}_{i}\!\left[z_{i,j}\right] - z_{k,j}  \label{eq:HITLMGA_weights_lombardi_delta}\\
w^{(k)}_{\text{HITL},j} &=
\begin{cases}
+1, & \delta_{k,j} > \tau, \\ 
0,  & |\delta_{k,j}| \le \tau, \\
-1, & \delta_{k,j} < -\tau, 
\end{cases} \label{eq:HITLMGA_weights_lombardi_discrete}\\
\mathbf{w}^{(k)}_{\text{HITL}} &= \big(w^{(k)}_{\text{HITL},1},\dots,w^{(k)}_{\text{HITL},q}\big)^\top\\
\mathbf{w}^{(\text{sum})}_{\text{\text{HITL}}} &= \sum_{k\in K} \mathbf{w}^{(k)}_{\text{\text{HITL}}} \\
j&\in \{1,\dots,q\}, \quad i \in I_{\text{MGA}} \notag
\end{align}
\end{subequations}

\subsection{Proposed Dynamic HITL-MGA Weights}
\label{sec:improvements_HITL_MGA}

The presented baseline HITL-MGA \cite{lombardiHumanintheloopMGAGenerate2025} has two key drawbacks: (i) reliance on a problem-specific distinctiveness threshold $\tau$ that must be retuned as operating scenarios (e.g., load) change, and (ii) discrete weights in $\mathbf{w}^{(k)}_{\text{HITL}}$ \eqref{eq:HITLMGA_weights_lombardi_discrete} that ignore how distinctive each variable in $\mathbf{z}$ actually is. Furthermore, in topology reconfiguration, another issue is redundancy. Solving \eqref{eq:basic_MGA} with different $\mathbf{w}_d$ can yield identical alternatives, an effect also seen in linear programs but more pronounced in sparse MILPs. The number of unique alternatives generated by a HITL-MGA approach serves as an indicator of its capability to explore the space of near-optimal topologies.

This paper addresses the shortcomings (i) and (ii) by modifying the encoding of the feedback in the weights as follows: we propose HITL-MGA-M, which takes the mean over all top-ranked alternatives instead of generating a set of weights $\mathbf{w}_{\text{HITL}}$ for each of them. In that way, HITL-MGA-M produces only one set of weights $\mathbf{w}_{\text{HITL}}$ that includes the feedback of all top-ranked alternatives:
\begin{equation}
    \label{eq:HITL_weights_V1}
    \delta_{j} = \mathbb{E}_i\!\left[z_{i,j}\right] - \mathbb{E}_k\!\left[z_{k,j}\right]
\end{equation}

HITL-MGA-M represents an intermediate step to HITL-MGA-MDy. Instead of using discrete weights, HITL-MGA-MDy scales the weights $\mathbf{w}_{\text{HITL}}$ dependent on the difference between the mean of the top alternatives and the mean of all alternatives. This renders the distinctive threshold $\tau$ obsolete and incorporates the degree of distinctiveness into the feedback:
\begin{equation}
\label{eq:dynamic_HITLMGA_weights}
     w_{\text{HITL},j} = \mathbb{E}_{i}[z_{i,j}] - \mathbb{E}_{k}[z_{k,j}] \\
\end{equation}

\begin{figure}[t]
  \centering
  \includegraphics[width=0.9\linewidth]{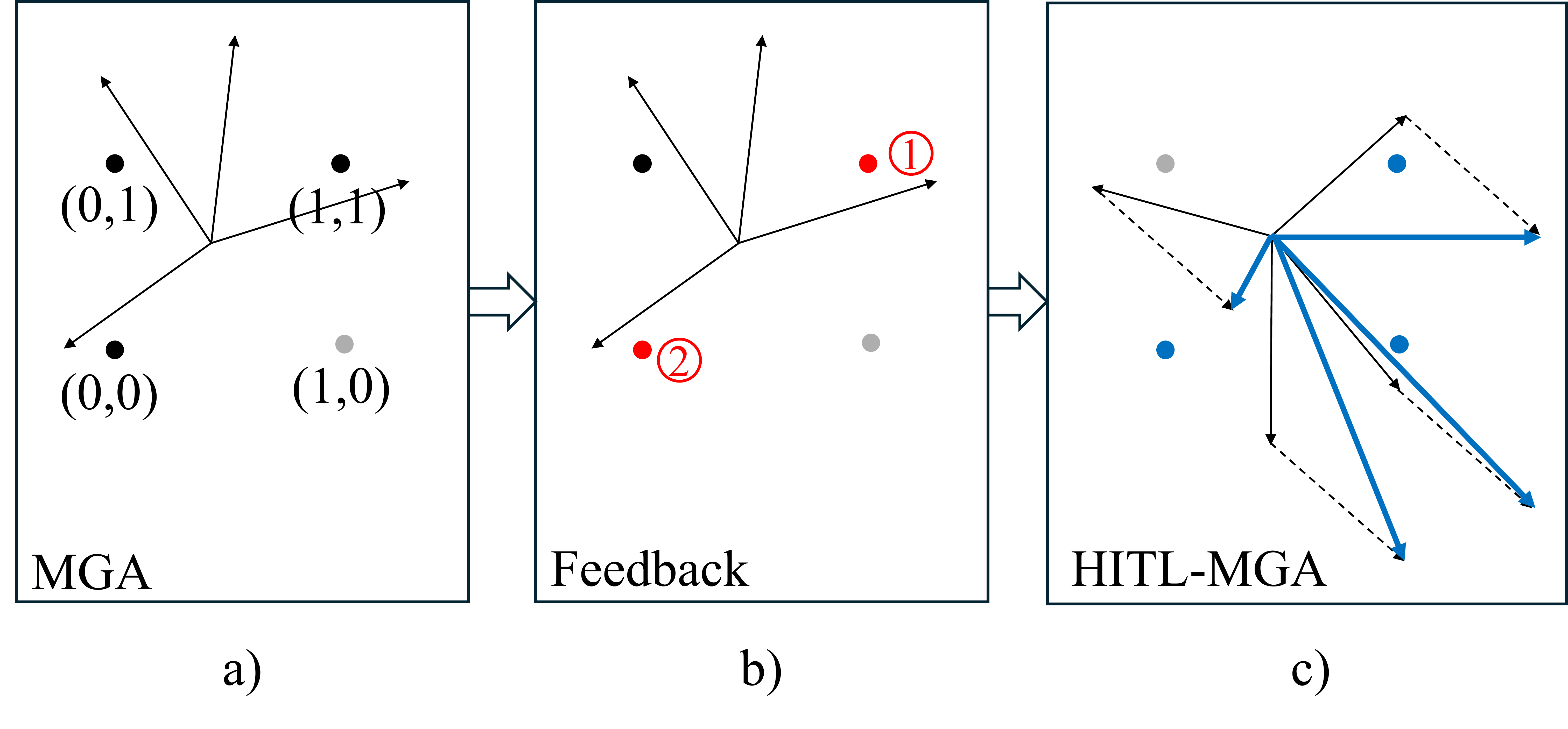}
  \caption{A two-dimensional representation of the proposed HITL-MGA-M \& HITL-MGA-MDy approach applied to binary decision variables. The dots represent topological alternatives in the near-optimal solution space. The black and blue arrows represent MGA objectives, and the dots in the respective colors represent the alternatives found by them.}
  \label{fig:HITL_MGA_vectors}
\end{figure}

Figure \ref{fig:HITL_MGA_vectors} visualizes how the adapted HITL-MGA approaches HITL-MGA-M and HITL-MGA-MDy approximate the black-box evaluation function from the information about the top-ranked alternatives with the linear function $B$ \eqref{eq:HITL_objective} using one-step learning. The figure shows a two-dimensional example.
Part a) shows the initial MGA. 
The feedback and the formation of the new sets of weights for the HITL-MGA objective are visualized in Figure \ref{fig:HITL_MGA_vectors} b) and c). 
The HITL-MGA approach encodes the feedback of the top-two-ranked alternatives into the weights $\mathbf{w}_{\text{HITL}}$, which, when combined with $\mathbf{w}_d$, result in new HITL-MGA objectives, represented by the blue arrows. The black arrows stand for $A$ \eqref{eq:HITL_objective}, which varies for each alternative. The dotted arrows represent $B$ \eqref{eq:HITL_objective}, which remains the same for all alternatives. The visualization does not consider the term $-\frac{s}{100 \cdot f^*}$, as it depends on the dispatch variables in $\mathbf{x}$ and substation reconfiguration variables in $\mathbf{y}$ that are not represented here.

\subsection{Proposed Feedback on Model Inaccuracies}
\label{sec:evaluation_functions}

Six exemplary evaluation functions $U$ are proposed to evaluate the HITL-MGA approaches, see Table \ref{tab:evaluation_functions}. 
Although some of the exemplary model inaccuracies could be removed by explicit modeling, we use them here to (i) produce the feedback for the ranking and (ii) quantify how well alternatives generated by different approaches address these inaccuracies.

\begin{table}[]
\centering
\caption{Exemplary evaluation functions and corresponding model inaccuracies.}
\label{tab:evaluation_functions}
\begin{tabular}{lll}
\hline
\textbf{Evaluation function}            & \textbf{Type of model inaccuracy}       & \textbf{Form}                                            \\ \hline
Specific   switching actions   & Missing expert knowledge       & $U_1(\mathbf{z})$                                  \\
Specific switching sets        & Missing expert knowledge       & $U_2(\mathbf{z})$                                 \\
Topological depth              & Missing expert preference      & $U_3(\mathbf{z})$                                 \\
Cumulative overload            & Missing expert preference      & $U_4(\mathbf{x},\mathbf{z})$ \\
Cumulative quadratic load            & Missing expert preference      & $U_5(\mathbf{x},\mathbf{z})$ \\
Switching sequence feasibility & \begin{tabular}[c]{@{}l@{}}Simplified modeling \\ assumption\end{tabular} & $U_6(\mathbf{x}, \mathbf{z})$
\end{tabular}%
\end{table}

The first three evaluation functions enable an assessment of to what extent the HITL-MGA approaches can incorporate feedback in the switching action space in the form of $U(\mathbf{z})$. In practice, this corresponds to a human operator aiming for specific switching actions ($U_1$) or sets of switching actions ($U_2$) to be part of the topology based on expert knowledge or preferences. The values of $U_1$ and $U_2$ should be maximized. For simplicity, the targeted switching actions ($J_{\text{spec}}$) and sets ($S_{\text{spec}}$) are determined by $f^*$. 
Another evaluation function is the topological depth. This aims at the expert preference to minimize topological depth, as formulated by \cite{viebahnC2GridOptionsTool} as the switching distance from the base topology $\mathbf{z}^{(0)}$. In this paper, the base topology considers all switches closed, i.e., $z_j = 1$:
\begin{subequations}
\label{eq:evaluation-functions_1to3}
\begin{align}
U_1(\mathbf{z}) 
&= \sum_{j \in J_{\text{spec}}} z_j,
\\
U_2(\mathbf{z}) 
&= \mathbb{I}\!\left\{\, z_j = 0 \;\; \forall\, j \in S_{\text{spec}} \right\},
\\
U_3(\mathbf{z}) 
&= \sum_{j \in J} \lvert z_j - z^{(0)}_j \rvert.
\end{align}
\end{subequations}

The other evaluation functions enable an assessment of to what extent the HITL-MGA approaches can incorporate feedback that also depends on other operational variables, i.e., dispatch: $U(\mathbf{x},\mathbf{z})$. This includes the cumulative overload ($U_4$) and cumulative quadratic load ($U_5$) over all lines $L$. The loads are determined by the power flow at each line $P_l$ and the line capacity $P_{\text{limit},l}$. Both evaluation functions should be minimized:
\begin{subequations}
    \begin{align}
        p_l &= \frac{P_l}{P_{\text{limit},l}}, \quad l \in L \\
        \delta_l &= p_l - 0.9 \\
        p_{\text{overload},l} &= 
        \begin{cases}
             \delta_l, \quad &\delta_l > 0 \\
             0, \quad &\delta_l \leq 0
        \end{cases}\\
        U_4(\mathbf{x},\mathbf{z}) &= \sum_{l\in L} p_{\text{overload},l} \label{eq:overload_evalfunction}\\
        U_5(\mathbf{x},\mathbf{z}) &= \sum_{l\in L} {p_l}^2 \label{eq:quadraticload_evalfunction}
    \end{align}
\end{subequations}

Finally, $U_6(\mathbf{x}, \mathbf{z})$ evaluates the feasibility of the line switching sequence under the alternative's dispatch. This addresses the simplified modeling assumption that we optimize topologies directly, while in reality, switches must be switched sequentially.

As the values of the cumulative overload, quadratic load, and the switching sequence feasibility depend on the dispatch and power flow variables in $\mathbf{x}$ and, therefore, indirectly also on the substation reconfiguration variables $\mathbf{y}$, the value of these evaluation functions for a given topology $\mathbf{z}$ is ambiguous. To remedy this issue, we assess a given topology by considering the dispatch and substation reconfiguration that minimizes the primary objective $f$. This is enforced by the buffer variable term $-\frac{s}{100\cdot f^*}$ used in the objective functions \eqref{eq:basic_MGA_objective} and \eqref{eq:HITL_objective}.

%% file: case_studies.tex
\section{Case Studies}
\label{sec:case_studies}
\begin{figure}[]
    \centering
    \includegraphics[width=0.7\linewidth]{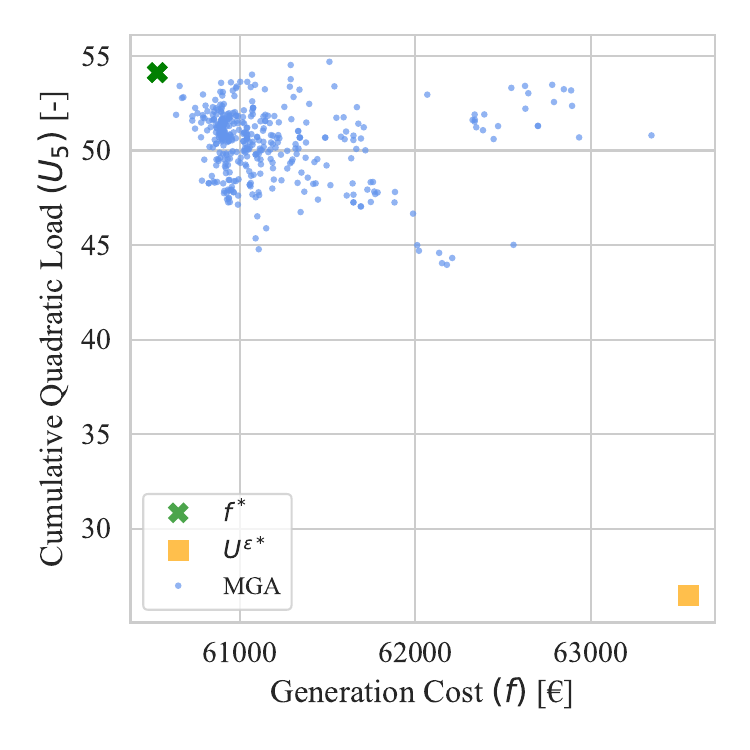}
    \caption{Performance of the MGA-generated alternatives with respect to the primary objective $f$ (generation cost) and to the evaluation function value of cumulative quadratic load ($U_5$) for the IEEE-118 bus system with line switching.}
    \label{fig:pareto_front}
\end{figure}

\begin{figure*}[]
\centering
    \subfloat[Evaluation function "specific switching actions" ($U_1$).  \label{fig:swaction_HITLMGA_comparison}]{         \includegraphics[width=0.32\textwidth]{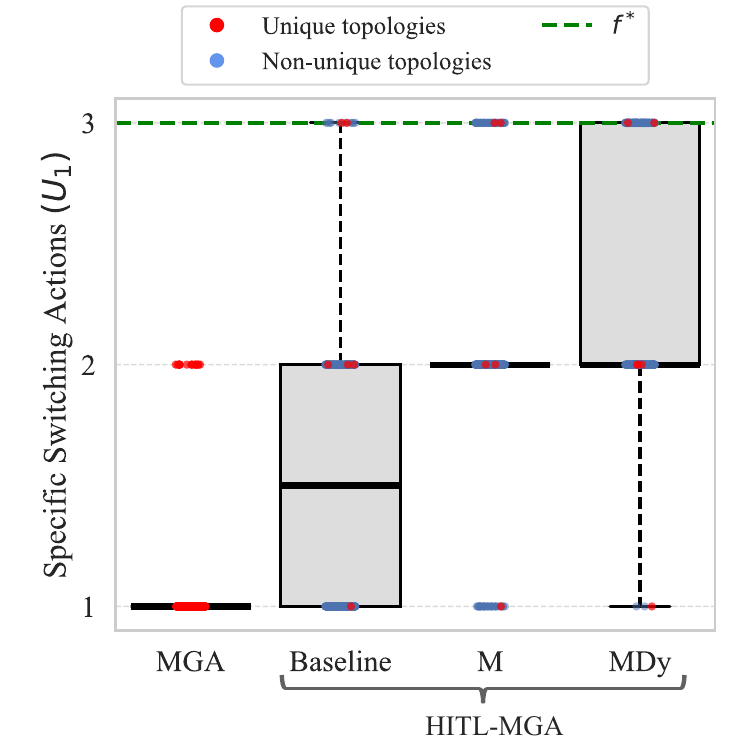}} \hfill                                               \subfloat[Evaluation function "cumulative overload" ($U_4$).\label{fig:cumoverload_HITLMGA_comparison}]{         \includegraphics[width=0.32\textwidth]{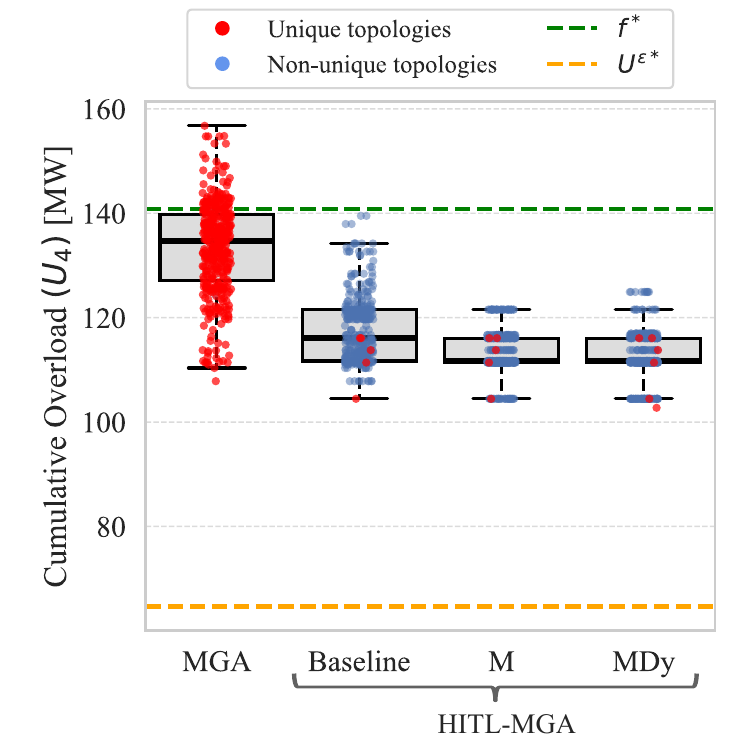}} \hfill   \subfloat[Evaluation function "cumulative quadratic load" ($U_5$).  \label{fig:cumquadraticload_HITLMGA_comparison}]{         \includegraphics[width=0.32\textwidth]{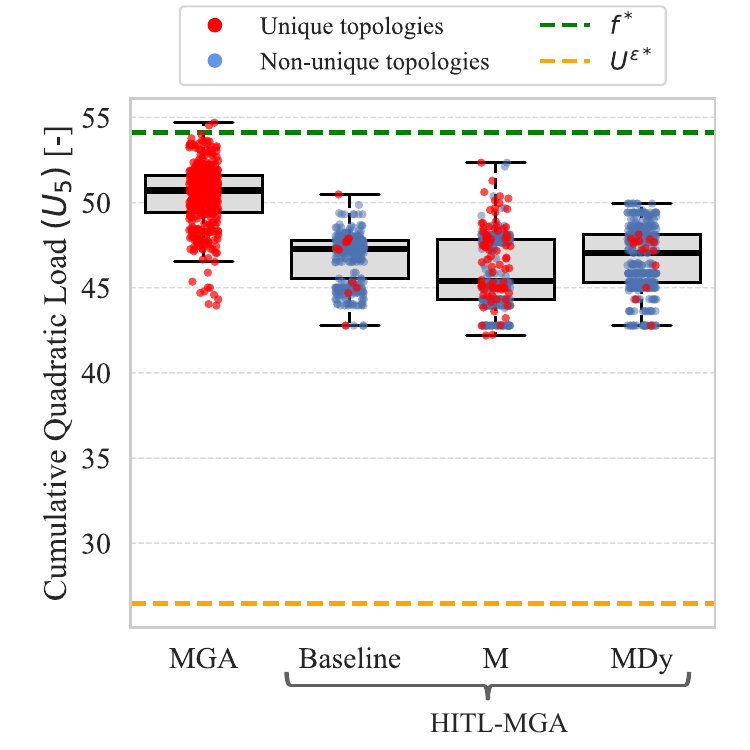}}
\caption{Performance comparison of the different HITL-MGA approaches (baseline, M, MDy) with respect to different model inaccuracies, evaluated by evaluation functions for the IEEE 118-bus system with line switching. For the specific switching actions, $f^* = U^{\varepsilon *}$. The aim is to maximize the number of specific switching actions, while the other evaluation functions should be minimized.}
\label{fig:HITLMGA_approach_comparison}
\end{figure*}

\subsection{Settings and Test Networks}
The case studies focus on IEEE 57-bus and 118-bus test networks \cite{babaeinejadsarookolaeePowerGridLibrary2021}. Congestion is enforced by reducing line capacities so that operating below rated limits requires topological actions. A line is considered overloaded above 90\% of its limit. 
We consider two topological action scenarios. In the base case, only line switching is possible. In the second case, covered in Section \ref{sec:substation_switching}, both line switching and substation switching are possible, with up to three line switching actions and three busbar splitting actions allowed. We only consider one snapshot of time with a fixed load for each test network. 
The electricity generation cost serves as the primary objective.
We assume $\varepsilon = 0.05$ and $\tau = 0.15$ \cite{lombardiHumanintheloopMGAGenerate2025}.

We generate 100 alternatives by sampling weights $\mathbf{w}_d$ and use the top 10 ranked alternatives for the HITL-MGA feedback. Using the feedback, 100 new alternatives are produced, using either the baseline HITL-MGA (Section \ref{sec:original_HITLMGA}), HITL-MGA-M, or HITL-MGA-MDy (both Section \ref{sec:improvements_HITL_MGA}). The performance of the different HITL-MGA methods is compared through the evaluation functions presented in Section \ref{sec:evaluation_functions}.  Each experiment is repeated with four random seeds. We also compute the optimum $U^{\varepsilon *}$ of each evaluation function within the $\varepsilon$-feasible set of \eqref{eq:basic_MGA_epsilon} to serve as a performance bound for the alternatives generated by MGA and HITL-MGA. The optimality MIP gap of the solver was set to $0.1\%$.

\subsection{Diverse Alternatives with MGA}
\label{sec:applying_MGA_to_NTR}

First, we investigate the performance of the alternatives generated by the basic MGA \eqref{eq:basic_MGA}. In all tested scenarios, MGA successfully generates topological alternatives within the near-optimality bound $\varepsilon$. Figure~\ref{fig:pareto_front} shows the IEEE 118-bus results with line switching for the cumulative quadratic load ($U_5$) \eqref{eq:quadraticload_evalfunction} evaluation function.
Most alternatives reach a better evaluation function value ($U_5$) compared to the least-cost topology $f^*$, while slightly trading off performance in the primary objective $f$. This behavior, where near-optimal solutions can perform better in unmodeled objectives than the optimal solution, is a core feature of MGA \cite{decarolisUsingModelingGenerate2011}. The performance of all generated alternatives is generally closer to the least-cost solution $f^*$ than to the evaluation function optimum $U^{\varepsilon *}$, due to the buffer variable term \eqref{eq:HITL_objective} and the resulting cost-optimal dispatch $\mathbf{x}$.

For other evaluation functions, such as topological distance ($U_3$) or switching sequence feasibility ($U_6$), the MGA finds few or no alternatives improving upon the least-cost solution $f^*$
(Table \ref{tab:HITLMGA_performance}). A possible reason for this is that some evaluation functions contradict the scope of the MGA objective formulation. For example, while the MGA objective aims at generating alternatives that are diverse in the switching action space, minimizing topological depth rewards low topological distance from the base topology.

\subsection{Reducing Model Inaccuracies with HITL-MGA}

Investigating the extent to which the HITL-MGA approaches guide the generation of alternatives, Figure \ref{fig:HITLMGA_approach_comparison} compares the performance of the different HITL-MGA approaches regarding three evaluation functions. 
When comparing the median results, the figure reveals that alternatives found by the HITL-MGA approaches outperform the basic MGA for the shown evaluation functions, while alternatives generated by HITL-MGA-M and HITL-MGA-MDy perform slightly better than the baseline approach. 
In Figure \ref{fig:cumoverload_HITLMGA_comparison} and \ref{fig:cumquadraticload_HITLMGA_comparison}, the performance of the HITL-MGA approaches appears to be loosely bounded by the initial MGA. 
This indicates that the exploratory part $A$ of the HITL-MGA objectives, with the chosen hyperparameters, is not sufficient to explore alternatives that perform drastically better than the top-ranked alternatives of the initial MGA. 
Furthermore, there is a significant performance gap for evaluation functions in Figure \ref{fig:cumoverload_HITLMGA_comparison} and \ref{fig:cumquadraticload_HITLMGA_comparison} comparing the evaluation function optimum $U^{\varepsilon *}$ and the HITL-MGA approaches. Two factors influence the size of this gap: (i) the lack of exploration in the initial MGA, and (ii) the cost-optimal dispatch variables $\mathbf{x}$ ensured by the buffer variable term, as observed in Section \ref{sec:applying_MGA_to_NTR}.

In Figure \ref{fig:HITLMGA_approach_comparison}, the color of the alternatives marks if its topology is unique, meaning that none of the alternatives generated either by MGA or the same HITL-MGA approach has the same topology $\mathbf{z}$. 
With the selected hyperparameters, the HITL-MGA approaches generate a significant fraction of redundant non-unique alternatives. This is likely encouraged by the small solution space due to the heavy congestion in the case studies. 
\begin{table}
\caption{Performance of HITL-MGA approaches for different evaluation functions for line switching.}
\label{tab:HITLMGA_performance}
\centering
\footnotesize
\setlength{\tabcolsep}{3pt}     
\renewcommand{\arraystretch}{1.1}
\label{tab:HITLMGA_performance}

\begin{tabularx}{\linewidth}{
  >{\raggedright\arraybackslash}p{9mm}  
  >{\raggedright\arraybackslash}p{35mm}       
  cccc}
\toprule
\textbf{System} & \textbf{Evaluation function}
& \textbf{\makecell{MGA\\valuable\\alternatives}}
& \multicolumn{3}{c}{\textbf{\makecell{HITL--MGA \\ more valuable \\ alternatives}}} \\
\cmidrule(lr){4-6}
& & & Baseline & M & MDy \\
\midrule
\multirow{6}{*}{\shortstack{IEEE\\57-bus}} 
& Specific switching actions        & $\checkmark$   & $\checkmark$ & $\checkmark$ & $\checkmark$ \\
& Specific switching sets           & $\checkmark$** & $\checkmark$* & --           & --           \\
& Topological depth                 & --             & --            & --           & --           \\
& Cumulative overload               & $\checkmark$   & $\checkmark$* & $\checkmark$* & $\checkmark$* \\
& Cumulative quadratic load         & $\checkmark$   & $\checkmark$  & $\checkmark$* & $\checkmark$ \\
& Switching sequence feasibility    & $\checkmark$   & $\checkmark$* & $\checkmark$* & $\checkmark$* \\
\midrule
\multirow{6}{*}{\shortstack{IEEE\\118-bus}} 
& Specific switching actions        & $\checkmark$   & $\checkmark$ & $\checkmark$ & $\checkmark$ \\
& Specific switching sets           & --             & --           & --           & --           \\
& Topological depth                 & --             & --           & --           & --           \\
& Cumulative overload               & $\checkmark$   & $\checkmark$ & $\checkmark$ & $\checkmark$ \\
& Cumulative quadratic load         & $\checkmark$   & $\checkmark$ & $\checkmark$ & $\checkmark$ \\
& Switching sequence feasibility    & $\checkmark$** & $\checkmark$* & --           & --           \\
\bottomrule
\end{tabularx}

\par\medskip\footnotesize * not unique \quad ** max. one alternative per random seed
\end{table}

\begin{figure*}[h]
\centering
    \subfloat[Performance of the HITL-MGA baseline with varying distinctive threshold. \label{fig:varying_distinctive_threshold}]{         \includegraphics[width=0.31\textwidth]{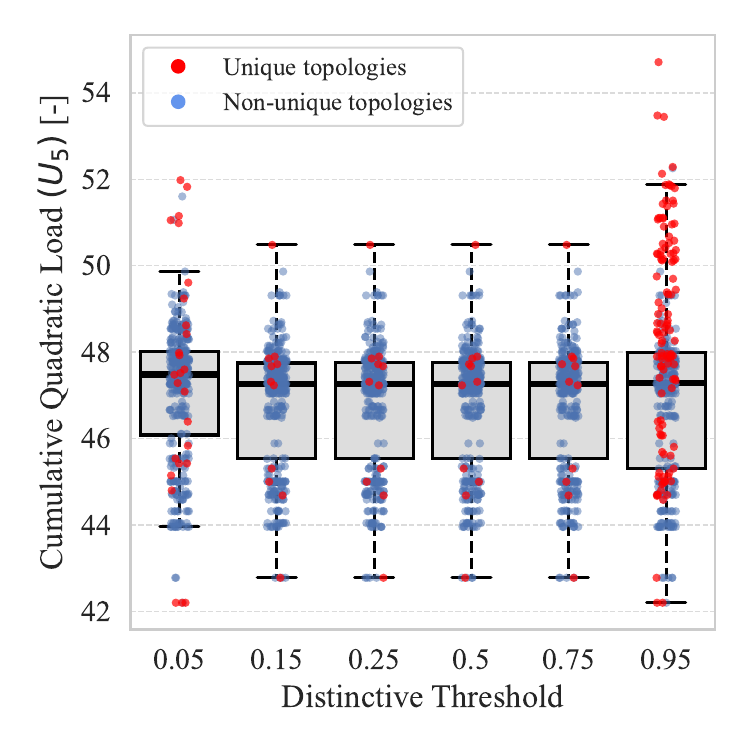}} \hfill                                               \subfloat[Performance of the HITL-MGA-MDy with varying $a/b$ ratio.\label{fig:varying_ab_ratio}]{         \includegraphics[width=0.31\textwidth]{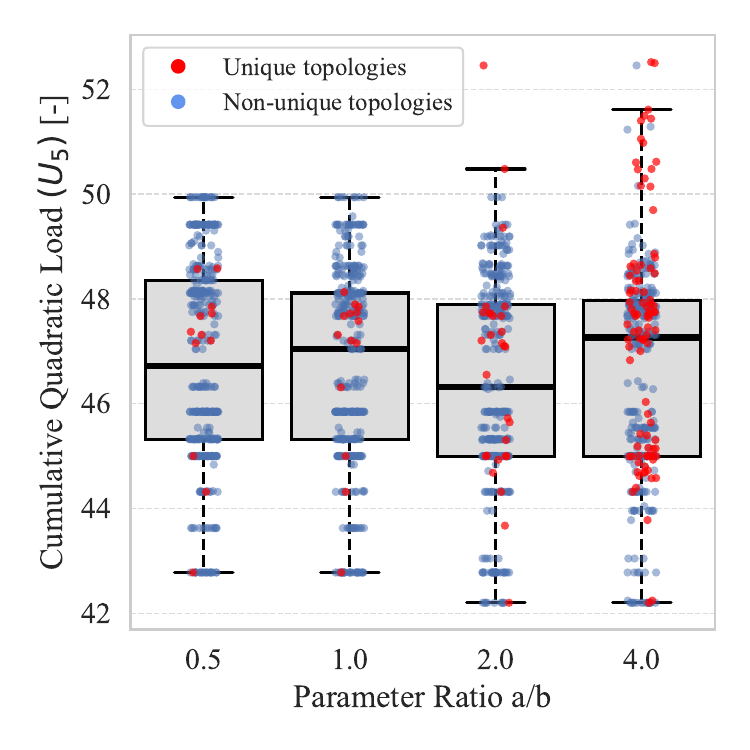}} \hfill   \subfloat[Performance of the HITL-MGA-MDy for 10 alternatives and 3 top-ranked alternatives.\label{fig:varying_number_alternatives}]{\includegraphics[width=0.31\textwidth]{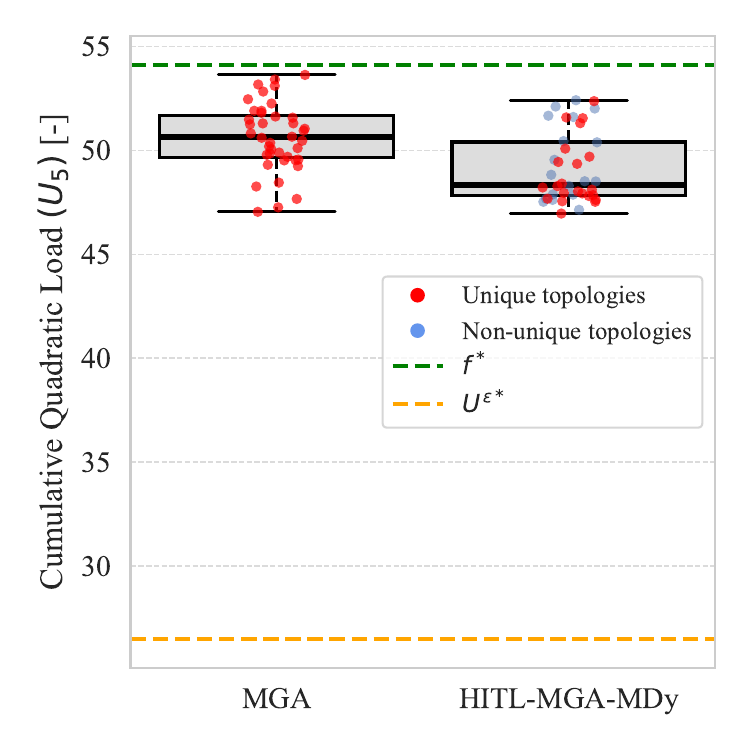}}
\caption{Impact of the hyperparameters on the HITL-MGA performance regarding minimizing the cumulative quadratic load ($U_5$) for the IEEE 118-bus system with line switching.}
\label{fig:hyperparameter_analysis}
\end{figure*}

\subsection{Different Model Inaccuracies}
Comparing the performance of the different HITL-MGA approaches for different evaluation functions, Table \ref{tab:HITLMGA_performance} summarizes the results across all model inaccuracies introduced in Section \ref{sec:evaluation_functions}. In the table, alternatives generated by MGA are considered "valuable" if not all of the alternatives have the same performance regarding the evaluation function. The alternatives of the HITL-MGA approach are considered "more valuable" when they contain a greater number of alternatives reaching at least the best performance level of MGA than the number of MGA alternatives achieving that best performance. 
For the evaluation functions of specific switching sets, topological depth, and switching sequence feasibility, the HITL-MGA approaches do not outperform the initial MGA for the 118-bus system.
When reducing the problem size to the IEEE 57-bus system, the HITL-MGA approaches perform better for the specific switching sets and the switching sequence feasibility, as the initial MGA finds more valuable alternatives here. However, as with the initial MGA approach, the HITL-MGA does not find valuable alternatives for topological depth.

\subsection{Sensitivity to the Hyperparameters}

Figure \ref{fig:hyperparameter_analysis} analyzes the impact of the different hyperparameters on the performance of the HITL-MGA approaches. As shown in Figure \ref{fig:varying_distinctive_threshold}, increasing $\tau$ from the proposed $15\%$ \cite{lombardiHumanintheloopMGAGenerate2025} to $75\%$ for the HITL-MGA baseline barely changes the outcomes. 
However, the alternatives generated for threshold values of $5\%$ and $95\%$ exhibit a larger spread and more unique alternatives. This behavior is due to the binary nature of the topology reconfiguration problem. 
As the mean of the switching variables over all MGA-generated alternatives is usually close to zero or close to one for the highly congested case studies, $\delta$ \eqref{eq:HITLMGA_weights_lombardi_delta} is also close to zero or close to one. 
Only if the threshold comes close to these values does it significantly influence the results. 

Another important hyperparameter is the ratio of the HITL-MGA objective parameters $a/b$. This ratio sets the trade-off between exploration and the exploitation of feedback in these approaches. Figure \ref{fig:varying_ab_ratio} shows that increasing the $a/b$ ratio yields more unique topologies and slightly better best alternatives, with similar medians.

When a human expert provides the HITL-MGA feedback, the number of alternatives should be limited to reduce complexity and time effort. 
Reducing the number of alternatives for the HITL-MGA-MDy approach to ten alternatives with three top-ranked alternatives slightly worsens median performance, shown in Figure \ref{fig:varying_number_alternatives}. Also, the best- and worst-performing alternatives that were found in the base case in Figure \ref{fig:cumquadraticload_HITLMGA_comparison} are lost. Nevertheless, the approach still yields valuable alternatives.

\subsection{Substation Switching}
\label{sec:substation_switching}
\begin{figure}[]
    \centering
    \includegraphics[width=0.7\linewidth]{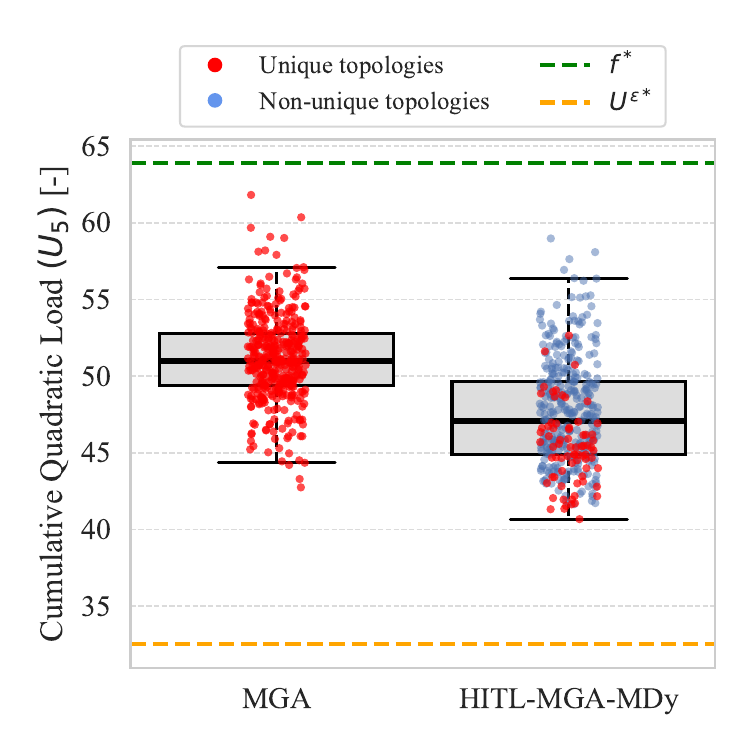}
    \caption{Performance of the HITL-MGA-MDy for substation switching regarding minimizing the cumulative quadratic load ($U_5$) for the IEEE 118-bus system.}
    \label{fig:substation_switching}
\end{figure}

As shown in Figure~\ref{fig:substation_switching}, applying HITL-MGA-MDy to line and substation switching yields higher median performance than the initial MGA. The larger action space also produces more diverse alternatives, indicating that HITL-MGA remains effective as dimensionality increases.

Comparing Figure \ref{fig:cumquadraticload_HITLMGA_comparison} and Figure \ref{fig:substation_switching}, the best HITL-MGA performance is better in the case of line and substation switching than for line switching only. Substation switching extends the feasible space of the topology reconfiguration problem, resulting in a lower optimum for the primary objective $f^*$. In turn, the allowed slack $\varepsilon f^*$ shrinks, tightening the $\varepsilon$-constraint, which may rule out near-optimal configurations that were found in the case with only line switching.

\section{Discussion}
\label{sec:discussion}
The presented HITL-MGA approach provides a structured way to generate diverse alternatives for topology reconfiguration that address specific model inaccuracies. HITL-MGA-MDy makes the parameter of the distinctive threshold $\tau$ redundant, while showing equal or better performance in incorporating feedback compared to the baseline. %and shows potential to improve further for continuous problems.
The HITL-MGA approaches can work for various model inaccuracies with evaluation functions in the form of $U(\mathbf{z})$ and $U(\mathbf{x,z})$. 

Some limitations remain. 
First, when the exploratory part of the MGA and HITL-MGA cannot explore good solutions for the evaluation function, e.g., for topological depth, HITL-MGA yields no valuable alternatives.
The success of the approach depends on the alignment between the evaluation function and the MGA or HITL-MGA objective, as well as the sampling of the weights $\mathbf{w}_d$. Adjusting the MGA objective, tuning the $a/b$-ratio, or improving the sampling of the weights $\mathbf{w}_d$ can help resolve these issues. 
Also, the performance of the proposed HITL-MGA is limited for evaluation functions in the form of $U(\mathbf{x,z})$, since variables $\mathbf{x}$ are optimized for the primary objective through the buffer variable term in the MGA objective.
Incorporating dispatch variables $\mathbf{x}$ and substation reconfiguration variables $\mathbf{y}$ (in case of substation switching) into MGA variables $\mathbf{z}$ and removing the buffer variable term could overcome this limitation, at the cost of higher dimensionality when sampling $\mathbf{w}_d$.

A second limitation arises under heavy congestion, where the feasible space is small and favors non-unique topologies. Increasing exploration via the $a/b$ ratio can mitigate this, without necessarily worsening median performance. Besides, larger action spaces (e.g., substation switching) reduce redundancy but increase exploration needs. This could also be resolved by improving the sampling of the weights $\mathbf{w}_d$.

%% file: conclusion.tex
\section{Conclusion}
\label{sec:conclusion}

This paper proposes a HITL-MGA approach for generating topological alternatives that address specific model inaccuracies in topology reconfiguration, covering both line and substation switching. The developed approach was investigated on the IEEE 57-bus and IEEE 118-bus systems. Three types of model inaccuracies were investigated through six evaluation functions, considering both switching variables and operational variables such as dispatch. The approach successfully generates valuable alternatives for all evaluation functions except topological depth, which was limited by the design of the MGA objective. The exploration capability of the initial MGA was identified as a bottleneck of the HITL-MGA approach, and the sampling of the weights $\mathbf{w}_d$ should be investigated further. In addition, future research could investigate alternative feedback mechanisms to enhance the effectiveness of the approach.